\documentclass[11pt]{article}
\usepackage{moriond,epsfig}

\def\Journal#1#2#3#4{{#1} {\bf #2}, #3 (#4)}
\def\Note#1#2#3#4{{#1} {\bf #2-#3-#4}}

\def\NPB{{\em Nucl. Phys.} B}
\def\PLB{{\em Phys. Lett.}  B}
\def\PRL{\em Phys. Rev. Lett.}

\def\JPG{{\em J. of Phys.} G : Nucl. Part. Phys.}
\def\JHEP{\em JHEP}

\def\ATNote{{\em ATLAS Note}}
\def\CMSNote{{\em CMS Note}}

\def\be{\begin{equation}}
\def\ee{\end{equation}}
\def\bea{\begin{eqnarray}}
\def\eea{\end{eqnarray}}

\begin{document}
\begin{flushright}
LPNHE-2002-05
\end{flushright}
\vspace*{4cm}
\title{Search for Extra Space-Dimensions at the LHC}

\author{ B. Laforge }

\address{Laboratoire de Physique Nucl\'eaire et de Hautes Energies\\
 CNRS-IN2P3 et Universit\'es Paris VI \& VII\\
 France}

\maketitle\abstracts{
\begin{center}
On behalf on the ATLAS and CMS collaborations
\end{center}
The introduction of extra space dimensions in the theory could be an elegant way to 
solve the hierarchy problem. There could even be one energy scale at 
which all interactions could unify. The limits coming from our knowledge of the 
gravitation at low distance allow this energy scale to be as low as few TeV. This 
situation is extremely interesting experimentally in the context of the LHC which 
will cover the range from 100 GeV to few TeV. This article describes the different 
analyses developed by the LHC experiments to study this new phenomenology.
}
\section{Motivations for Searching for Extra Space Dimensions}
The Standard Model of particle physics has a very powerful prediction power but 
provides no answer to very important conceptual questions related to the deep 
structure of our universe. A part of these questions concern the relations 
that exist between the different interactions that we know. Are these forces 
originating from a unique force which would lead to a Grand Unification 
Theory (GUT) ? What is exactly the Plank Scale ? Is there a common framework in 
which all the interactions could be embedded in a self-consistent theory ? What 
is the exact structure of our space-time and how does it interfere with the 
particle spectrum that we observe ? Why do various energy scales exist in 
physics going from 200~GeV for the electroweak scale to $10^{19}$~GeV for the 
Plank scale ?  

In the last past years, a new direction\cite{Anto,ABQ,ADD,RS} has been proposed to 
explain why gravitation is so weak and why the Plank scale is so large. The 
new idea is to introduce extra-space dimensions on top of our usual 3 dimensions. 
These extra-dimensions are not seen at our scale, so they have to be compactified 
with a size smaller than the one already experimentally investigated. This size 
is very small for all the electroweak and strong interactions but is of the order 
of $0.15$~mm for the gravitation\cite{gravlim}. So there is a possibility that 
these hypothetical extra-dimensions, where gravity could propagate, are quite 
large compactified dimensions. As a result of this situation, the usual Plank 
mass could be an irrelevant physical scale whereas the true gravitational 
scale could be much smaller, as small as few TeV. 

Introducing such extra space dimensions could then solve the energy scale hierarchy 
or, at least, reformulate it into a geometric problem. This new framework is 
motivated theoretically by the fact that the only framework in which we could 
treat on the same bases all the interactions, including gravitation, is the string 
theory framework which needs extra-space dimensions for self-coherence reasons. So 
the question is~: can we be experimentally sensitive to these hypothetical 
extra-dimensions ?  
\section{Strategy of the Searches}
The main characteristics of the models of extra-dimensions is that the gravitational 
interaction could become non-negligible at small distances. Gravitation is brought in 
the field of particle physics experimentation through coupling of graviton to the usual 
known particles. In these models, on each vertex or on each 
propagator of a given model (SM, MSSM,...), one can plug a graviton leg. 
Having all these new vertices and the related Feynman rules, one can construct 
all the associated Feynman graphs for a given final state. In these models, 
our matter fields are bound on a 4D hypersurface of the whole (4+n)D universe 
while the graviton can go into the extra-dimensions (the bulk). The phenomenology 
is enriched when the usual gauge bosons can also propagate in the extra-dimensions. 
Boundary conditions that exists in the extra dimensions imply that their 
momentum component in the extra dimension is quantized. In our 4D world, this 
would be seen as the appearance of a infinite number of massive states. These 
so called Kaluza-Klein (KK) states are a clear experimental signal that could 
be observed experimentally either through their direct production or their 
virtual exchange contribution to standard cross-sections. Another effect of 
these KK states could be a modification of the renormalisation group equations, 
leading to possible large modifications of cross-sections. Beyond the interplay 
between the SM and the introduction of extra-dimensions for which the new 
phenomenology is already very exciting, one can also consider the modification 
of other, in particular SUSY, models. The LHC collaborations have studied 
the main models introducing extra-space dimensions. The two classes of models 
differ in the way they implement the space-time metric structure.
The model implementing large extra-dimensions proposed by N.~Arkhani-Hamed, 
S.~Dimopoulos and G.~Dvali\cite{ADD} has a factorizable metric - the metric 
has a diagonal block structure - whereas in the Randall-Sundrum\cite{RS} there 
is an interplay between the (3+1)D metric and the value of the extra-dimensional 
coordinate. In this latter model, the dependence is exponential and is responsible 
for the shrinkage of the strength of the gravitational interaction in our (3+1)D 
subspace, often called a (3+1)-brane.
\subsection{Direct Production of Gravitons}
\begin{figure}
\rule{5cm}{0.2mm}\hfill\rule{5cm}{0.2mm}
\begin{center}
\epsfig{figure=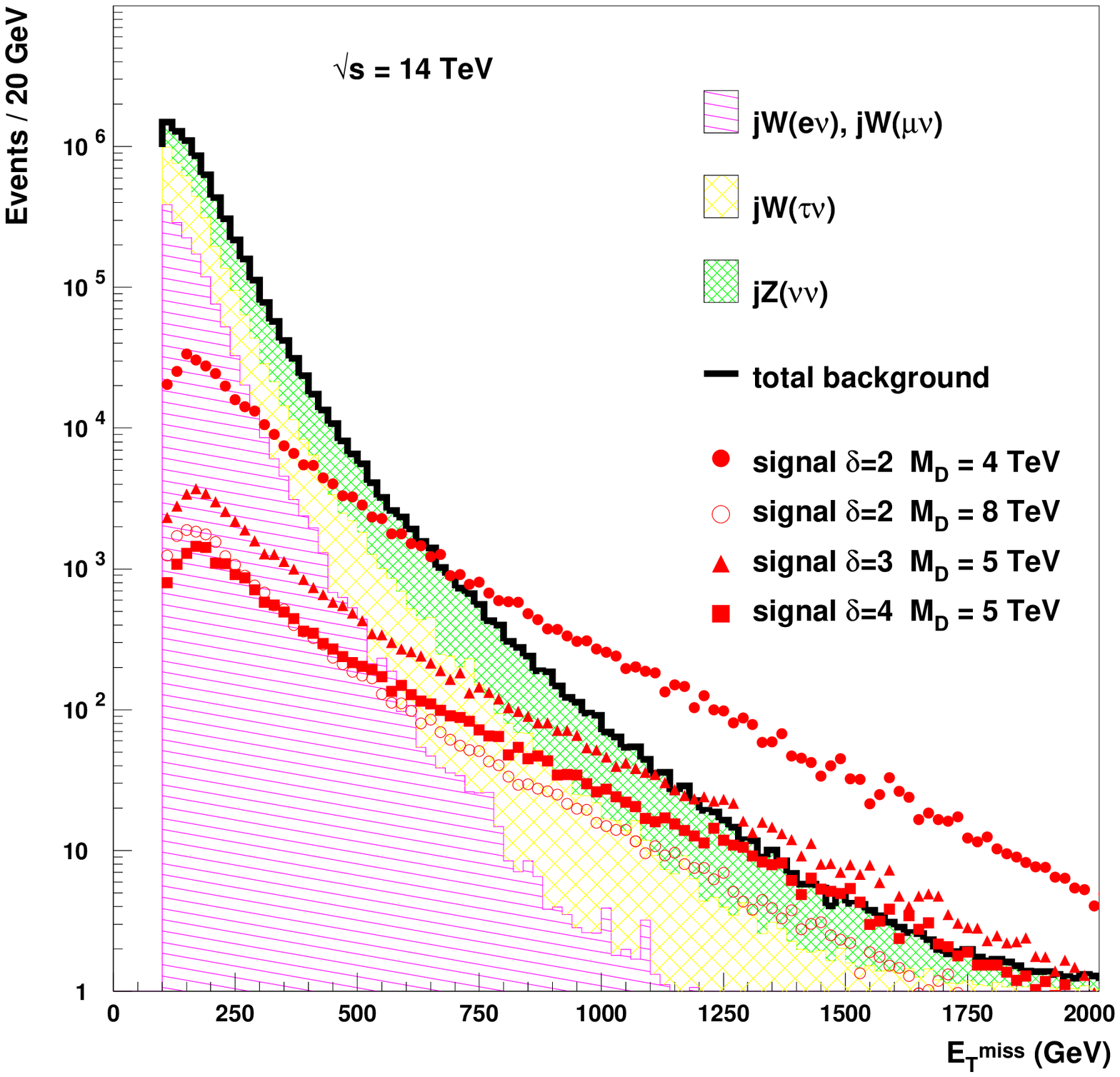,height=10cm}
\end{center}
\rule{5cm}{0.2mm}\hfill\rule{5cm}{0.2mm}
\caption{
\label{fig:VI}}
\end{figure}

Direct production of graviton would constitute a very nice signal at the LHC. Indeed, 
in the simplest channels, this graviton would be produced together with a gluon, 
a Z boson or a photon. As the graviton interacts only gravitationally and extra 
dimensions are open for it, it will not interact in our detector, giving 
rise to a missing transverse momentum in the event. This missing $E_t$ can be 
very large, much beyond 1 TeV.  ATLAS has studied the graviton + jet channel\cite{VI}
for which the dominant channel is $qg \longrightarrow gG$. The signal cross-section 
has been implemented in ISAJET using an effective theoretical approach. The generated 
events have been investigated using the ATLAS fast simulation program\cite{ATLFAST}. 
Figure \ref{fig:VI} shows the signal shapes as detected in ATLAS as a function of 
the missing transverse energy for various cases of large extra-dimension model 
parameters and for SM distributions. This study shows that after one year of high 
LHC luminosity, ATLAS will be able to discover a fundamental scale up 9 to 6 TeV 
for 2 or 4 extra dimensions, respectively.
\subsection{Virtual Exchange of Gravitons}
\begin{figure}
\rule{5cm}{0.2mm}\hfill\rule{5cm}{0.2mm}
%\vskip 2.5cm
\begin{minipage}[b]{0.5\linewidth} % A minipage that covers half the page
\centering
\epsfig{figure=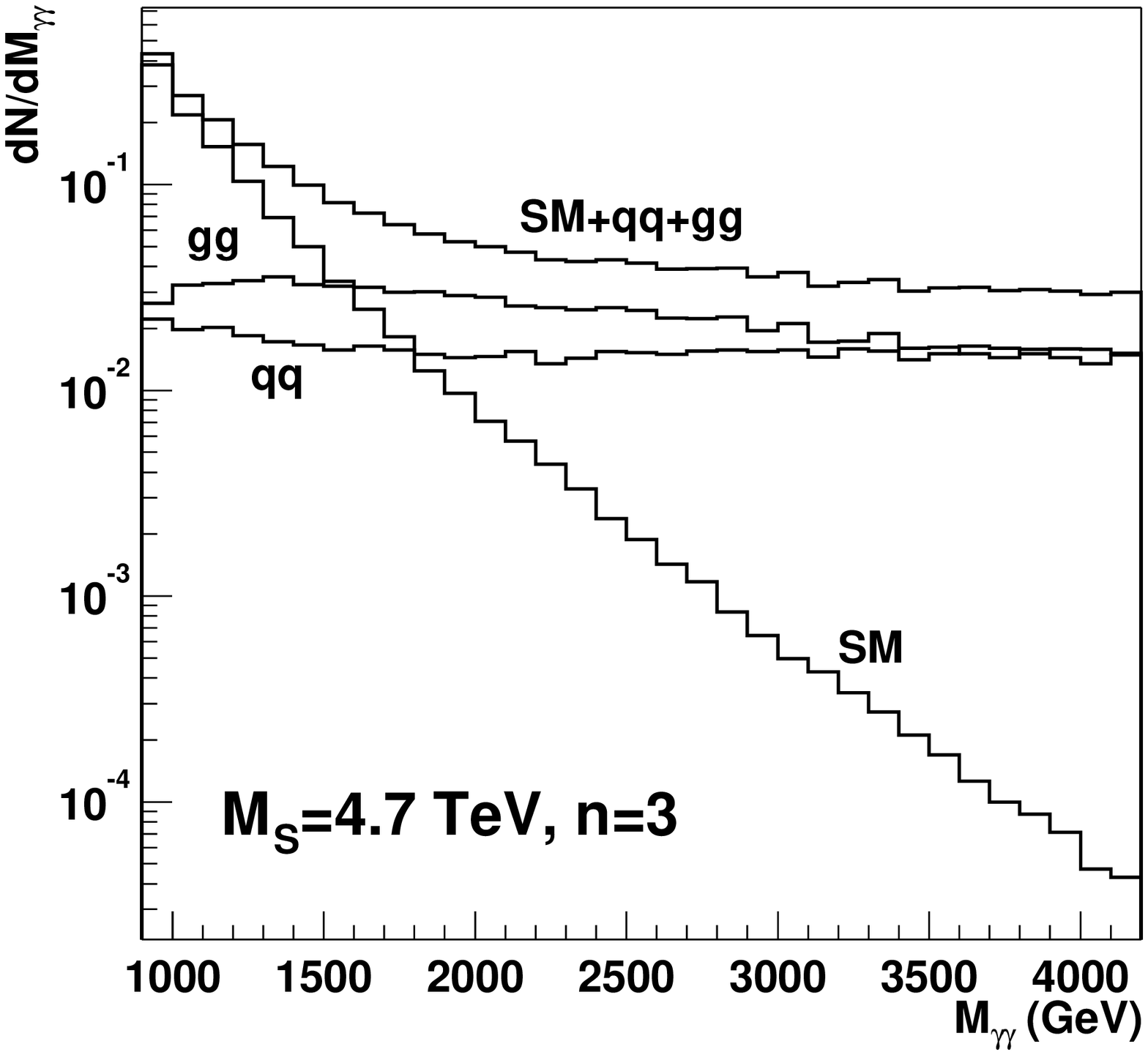,width=6cm}
\caption{{\small $pp \longrightarrow \gamma \gamma$ cross-section vs di-$\gamma$ 
invariant mass for SM model and for 3 extra-dimensions and $M_s = 4.7$~TeV. The 
two contributions labeled $q\bar{q}$ or $gg$ correspond to a graviton exchange 
with a $q\bar{q}$ or $gg$ initial state.}}
\label{fig:KMZ}
\end{minipage}
\hspace{0.5cm} % To get a little bit of space between the figures
\begin{minipage}[b]{0.5\linewidth}
\centering
\epsfig{figure=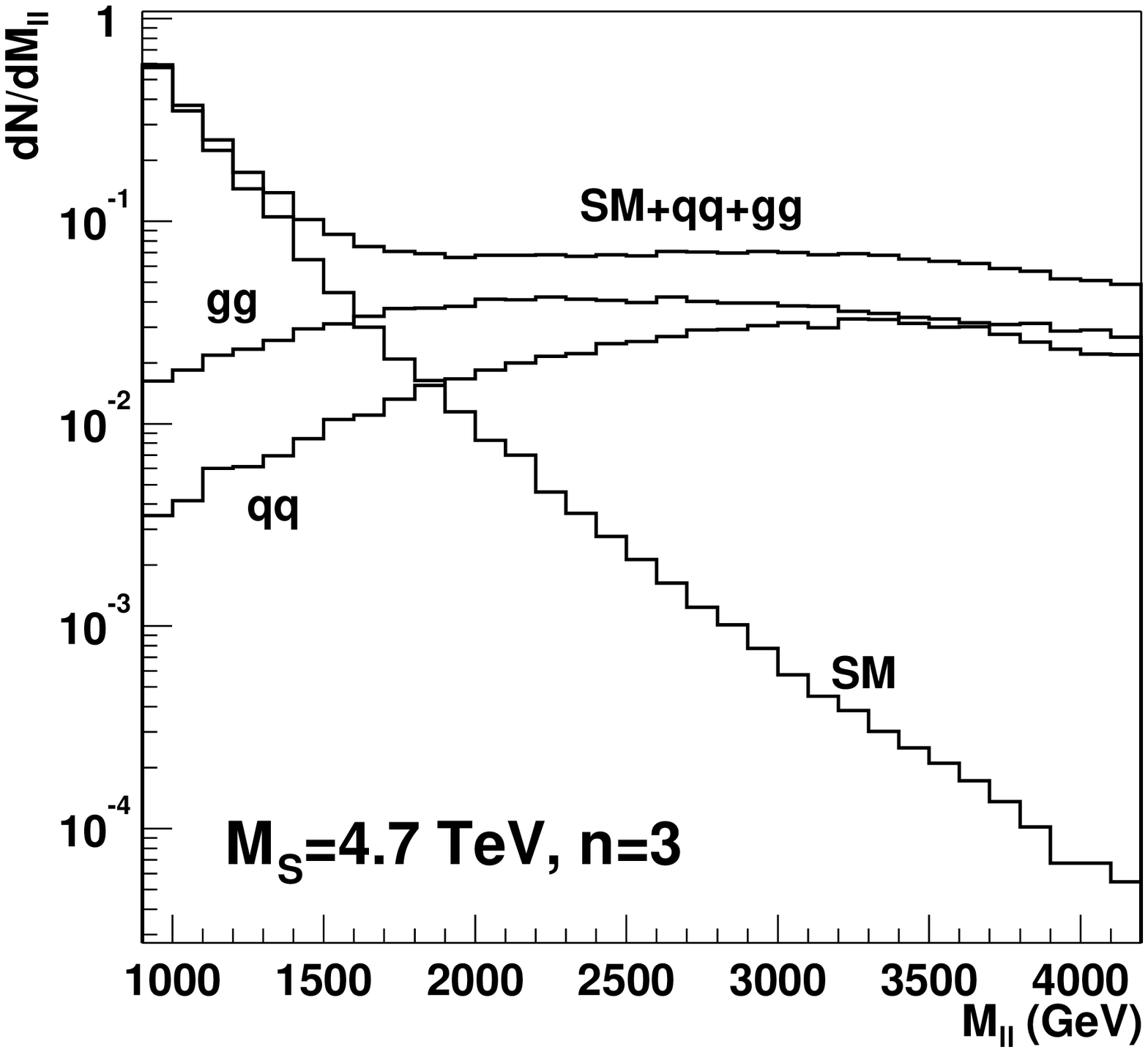,width=6cm}
\caption{$pp \longrightarrow l^+ l^-$ cross-section versus the di-lepton invariant 
mass for SM model and for 3 extra-dimensions and $M_s = 4.7$~TeV. The two 
contributions labeled $q\bar{q}$ or $gg$ correspond to a graviton exchange with 
a $q\bar{q}$ or $gg$ initial state.}
\end{minipage}
\rule{5cm}{0.2mm}\hfill\rule{5cm}{0.2mm}
\end{figure}

\begin{figure}
\rule{5cm}{0.2mm}\hfill\rule{5cm}{0.2mm}
\begin{center}
\includegraphics*[width=8cm]{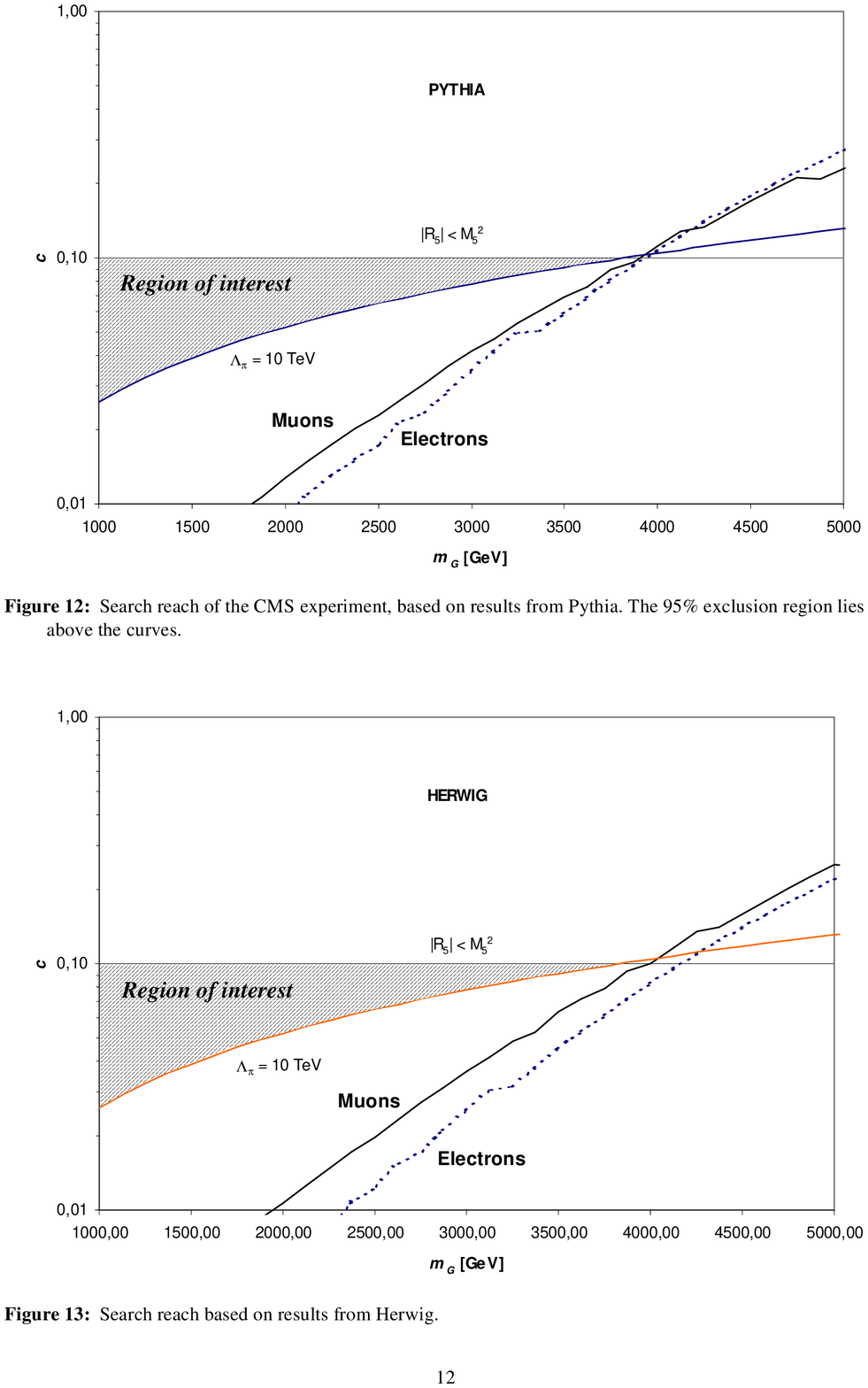}
\caption{CMS will be able to reach on the existence of a graviton in the 
Randall-Sundrum model using one of the two leptonic channels $e^+e^-$ and 
$\mu^+ \mu^-$
\label{CMS:ROI}}
\end{center}
\rule{5cm}{0.2mm}\hfill\rule{5cm}{0.2mm}
\end{figure}

\begin{figure}
\begin{center}
\rule{5cm}{0.2mm}\hfill\rule{5cm}{0.2mm}
\vskip .3cm
\epsfig{figure=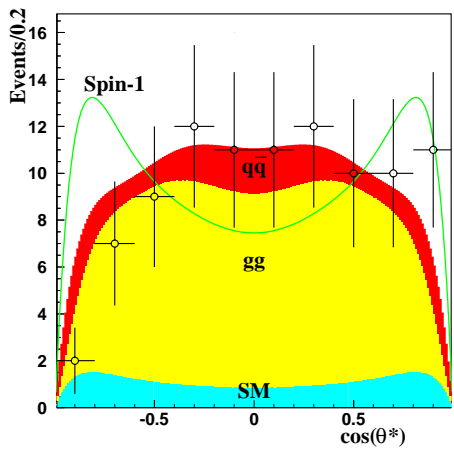,height=6cm}
\caption{Center of mass frame polar angle distribution of final state electron pair
direction is shown for a graviton of 1.5 TeV and for one year of LHC data at high 
luminosity ($100~fb^-1$). Solid line shows the expected signal shape for a spin-1 
resonance.
\label{fig:spin2}}
\rule{5cm}{0.2mm}\hfill\rule{5cm}{0.2mm}
\end{center}
\end{figure}

As a quark and an antiquark annihilate into a virtual $\gamma$ or Z, they can 
also produce a virtual graviton (or any of its KK modes). The multiplicity of 
KK states can give a large contribution to the production cross-section of any 
final state such as di-jet, di-lepton, etc. The sum of contributions to be taken 
into account is divergent as soon as there is more than one extra-dimension. So, 
in a quantum field theory framework, it has to be regularized giving rise to a 
low energy effective theory. This regularisation, usually done using an energy 
cut $\Lambda$,makes difficult to understand the exact relationship between 
the observable signals and the related number and scale of extra dimensions. 
Nevertheless, both ATLAS and CMS collaborations have studied this kind of signals, 
trying to understand the sensitivity of their detectors to this phenomenology. 
The first ATLAS study\cite{KMZ} focused on the channels  $pp \longrightarrow l^+ 
l^- +X$ and $pp \longrightarrow \gamma \gamma + X$. Figure \ref{fig:KMZ} 
shows the signal shape for the two channels as a function of the two final state 
particles invariant mass. For a luminosity of 100~$fb^{-1}$ (one year of LHC at 
high luminosity),  a sensitivity to an energy scale of 7 to 8 TeV is reached for 
a number of extra dimensions varying between 2 and 5.
A similar analysis in the di-lepton channel has been made in CMS but in the context 
of the Randall-Sundrum model\cite{LPFVLW}. Phenomenological constraints on this 
model\cite{EtudePhenoRS} defines a ``region of interest'' in the ($M_{graviton}, coupling$) 
plane, shown in  Figure \ref{CMS:ROI}. This figure also presents the limits that can 
be reached using the CMS detector both for the electronic or muonic channels. The conclusion 
of that study is that CMS will be able to cover the whole region of interest.
The ATLAS collaboration has also developed a model independent search of narrow 
gravitons\cite{AOPW} studying the di-electron production channel 
$gg(q\bar{q})\longrightarrow Graviton \longrightarrow e^+e^-$. 
Such a resonance could be seen in the range $2-5$~TeV in the case of the 
Randall-Sundrum model in the case when the coupling is the smallest allowed 
by the phenomenological constraints ($k/\Lambda_{\pi}>0.01$)\cite{EtudePhenoRS}. 
Angular distribution of polar angle of the electrons with respect to
the direction of the center of mass of the pair has been performed to
confirm the spin-2 property of the graviton. ATLAS will be able to distinguish a 
spin 2 resonance form a spin-1 resonance up to the mass of 1.7 TeV. Figure 
\ref{fig:spin2} shows the angular distribution expected in ATLAS for a spin-2 
graviton of 1.5~TeV mass and for an integrated luminosity of 100~$fb^{-1}$ 
corresponding to one year of data taking at LHC high luminosity. A solid line 
shows what would be the signal shape for a spin-1 resonance.
ATLAS also studied the possibility to observe KK modes of the SM gauge bosons\cite{AP} 
if the compactification scale is of the order of few TeV. Even in absence of KK 
resonances, a careful study of the di-lepton cross-section can lead to a 
determination of the compactification scale. 

\subsection{Radion production in the Randall Sundrum Model}

In the case of the Randall Sundrum Model, the space time metric structure is 
modified in a non factorizable way and we end up with the following metric~:
$$
ds^2 = e^{-2 {\bf k} r_c |y|}\; \eta_{\mu \nu} dx^{\mu} dx^{\nu} -dy^2
$$
where $r_c$ is the compactification radius and {\bf k} is a number that reflects 
the expectation value of a field called the graviscalar Radion. A mechanism 
introduced by Goldberger and Wise\cite{GoldbergerWise} uses this field to 
stabilize $r_c$. The radion can be massive but could likely be less massive 
than the first Kaluza-Klein state of the graviton. The phenomenology of such 
a scalar is similar to that of the Higgs boson, and its existence could be 
confirmed by the evaluation of the decay branching ratios of the scalar field. 
Based on studies on the SM Higgs, limits on the observability of the radion 
could be obtained\cite{ACVP}. ATLAS has also investigated the situation where 
the radion would be heavier than twice the Higgs mass and studied its decay 
in the 2 Higgs channel for two different final states of the Higgs decay: 
$bb \gamma\gamma$ and $bb \tau \tau$. 
 
\subsection{Effects on the Renormalisation Group Equations}

\begin{figure}
\rule{5cm}{0.2mm}\hfill\rule{5cm}{0.2mm}
%\vskip 2.5cm
\begin{minipage}[b]{0.5\linewidth} % A minipage that covers half the page
\centering
\epsfig{figure=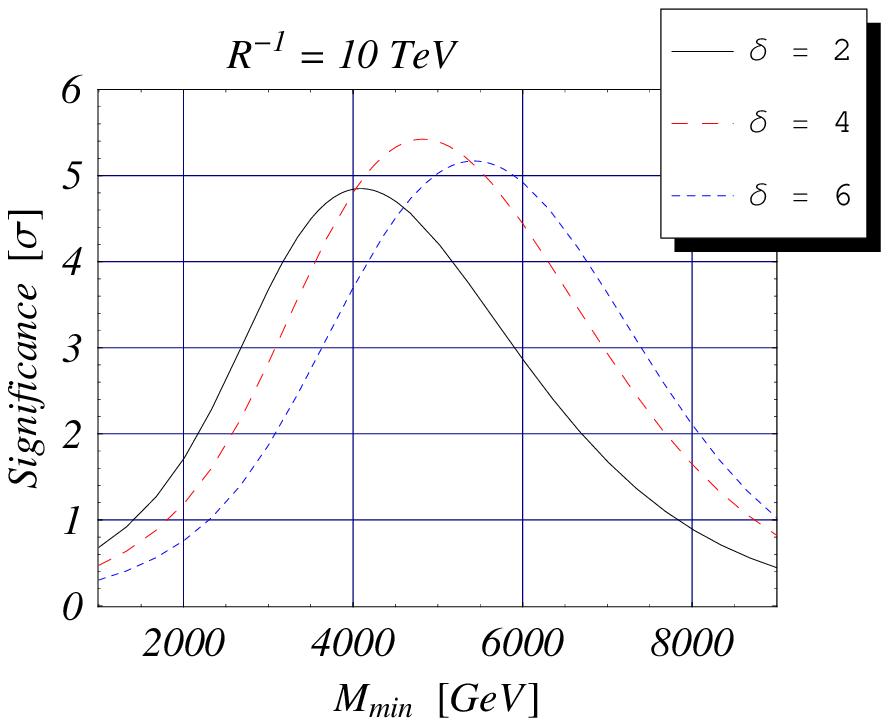,width=8cm}
\caption{{\small The statistical significance (in units of $\sigma$'s)
             of the deviation from the Standard Model,
             as the function of the minimal dijet mass $M_{min}$ at the LHC, 
             for different numbers of dimensions and for a compactification 
             scale of 10 TeV.}}
\label{FigSignificance1}
\end{minipage}
\hspace{0.5cm} % To get a little bit of space between the figures
\begin{minipage}[b]{0.5\linewidth}
\centering
\epsfig{figure=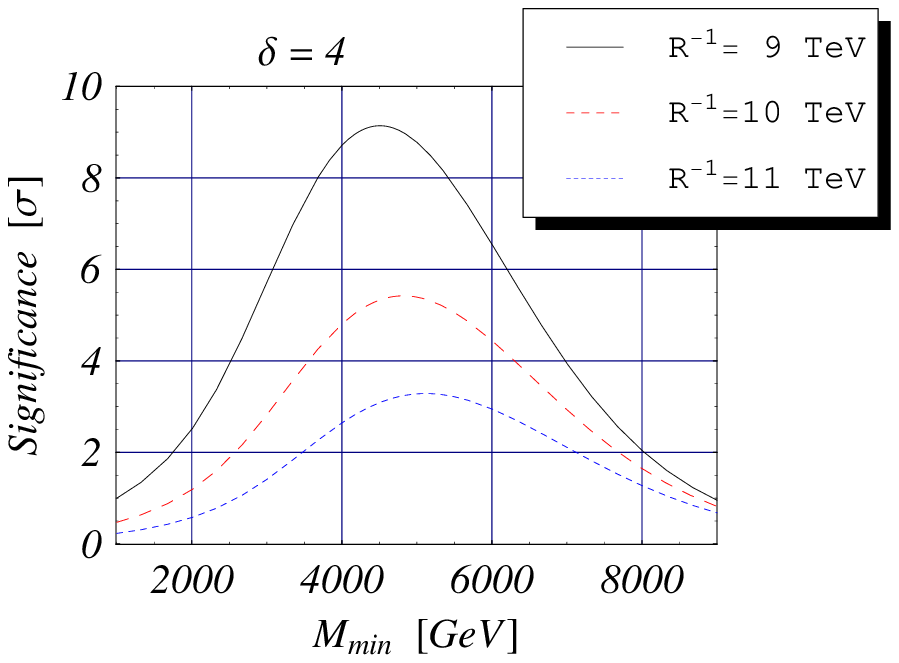,width=8cm}
\caption{\small{The statistical significance (in units of $\sigma$'s)
             of the deviation from the Standard Model,
             as the function of the minimal dijet mass $M_{min}$ at the LHC, 
             for different compactification scales and for 4 extra dimensions. }}
\label{FigSignificance2}
\end{minipage}
\rule{5cm}{0.2mm}\hfill\rule{5cm}{0.2mm}
\end{figure}

It can also happen that the extra-dimensions are open to the standard gauge bosons. 
In that case, it has been shown\cite{DDG} that the renormalisation group 
equations can be modified such that the coupling evolution will be strongly altered 
from the standard logarithmic behavior. In such a case, unification could be 
obtained at a low energy scale down to few tens of TeV. ATLAS studied\cite{BL} 
this possibility looking at a possible suppression of di-jets production as 
$\alpha_s$ decreases quickly with the energy scale. This study was done in the 
framework of a large extra-dimension model. A compactification scale from 5 to 10~TeV
could be reached using one year of high luminosity LHC data. 
Figures \ref{FigSignificance1} and \ref{FigSignificance2} show the significance 
that can be achieved in the dijet channel in ATLAS for $100\;pb^{-1}$ and for 
different radii of the compactification scale or for different number of extra 
dimensions.
  
\subsection{Extra Dimensions and Supersymmetry}

The existence of extra-dimensions could also have unexpected consequences. As an 
example, ATLAS has investigated the possibility to discover supersymmetry through 
the charged Higgs decay to a $\tau$ and a neutrino\cite{AD}. In the usual MSSM, 
the decay into a neutrino with a right helicity is strongly suppressed but this 
could be changed if the right-handed neutrino can go in the bulk. Indeed,
conservation laws do not force the right-handed neutrino to our brane. In case 
where these neutrinos could go to the bulk then the phase space available 
for their production would enhance the charged Higgs decay width in the suppressed 
channel. The ATLAS analysis shows that this channel is a very good candidate to 
search for both extra-dimensions and SUSY.

\section{Conclusion}

The ATLAS and CMS collaborations have studied intensively the different models involving 
extra-space dimensions.It appears that the LHC experiments will be very sensitive to 
the related signals and should be able to either discover these dimensions or draw 
limits to a possible compactification scale above 5 to 10 TeV, depending on the 
models.

\section*{Acknowledgments}

I take the occasion here to thank the ATLAS and CMS members who helped me a lot preparing my talk. I especially 
want to thank G. Azuelos, F. Gianotti, K. Jacobs, L. Poggioli for their valuable comments and for the reading 
of that manuscript.

\end{document}